\documentclass{IEEEtran4PSCC}
\usepackage{cite}
\usepackage{amssymb,amsfonts}
\usepackage{algorithmic}
\usepackage{graphicx}
\usepackage{textcomp}
\usepackage{xcolor}
\usepackage{subcaption}
\usepackage[cmex10]{amsmath}
\makeatletter
\let\old@ps@headings\ps@headings
\let\old@ps@IEEEtitlepagestyle\ps@IEEEtitlepagestyle
\def\psccfooter#1{%
    \def\ps@headings{%
        \old@ps@headings%
        \def\@oddfoot{\strut\hfill#1\hfill\strut}%
        \def\@evenfoot{\strut\hfill#1\hfill\strut}%
    }%
    \def\ps@IEEEtitlepagestyle{%
        \old@ps@IEEEtitlepagestyle%
        \def\@oddfoot{\strut\hfill#1\hfill\strut}%
        \def\@evenfoot{\strut\hfill#1\hfill\strut}%
    }%
    \ps@headings%
}
\makeatother

\begin{document}
%
\title{Small-Signal Stability and SCR Enhancement of Offshore WPPs with Synchronous Condensers}
\author{
    \IEEEauthorblockN{Sulav Ghimire\IEEEauthorrefmark{1}\IEEEauthorrefmark{2}\IEEEauthorrefmark{3}, Kanakesh V. Kkuni\IEEEauthorrefmark{1}, Emerson D. Guest\IEEEauthorrefmark{1}, Kim H. Jensen\IEEEauthorrefmark{1}, Guangya Yang\IEEEauthorrefmark{2}}
    \IEEEauthorblockA{\IEEEauthorrefmark{1}\textit{Siemens Gamesa Renewable Energy A/S}, 7330 Brande, Denmark}
    \IEEEauthorblockA{\IEEEauthorrefmark{2}\textit{Technical University of Denmark}, 2800 Kgs Lyngby, Denmark}
    \IEEEauthorblockA{\IEEEauthorrefmark{3}Correspondence via: \textit{sulav.ghimire@siemensgamesa.com}}
}


\maketitle


\begin{abstract}
Synchronous condensers (SCs) have been reported to improve the overall stability and short-circuit power of a power system. SCs are also being integrated into offshore wind power plants (WPPs) for the same reason. This paper, investigates the effect of synchronous condensers on an offshore wind power plant with grid-following (GFL) and grid-forming (GFM) converter controls. Primarily, the effect of synchronous condensers can be two-fold: (1) overall stability enhancement of the WPP by providing reactive power support, (2) contribution to the effective short circuit ratio (SCR) of the WPP by fault current support. Therefore, this paper focuses on studies concerning these effects on an aggregated model of a WPP connected to the grid. To that end, a state-space model of the test system is developed for small-signal stability assessment and the synchronous condenser's effect on its stability. In addition, a mathematical explanation of SCR enhancement with synchronous condenser is provided and is verified with time-domain electromagnetic transient simulations.
\end{abstract}



\section{Introduction}
Modern offshore wind power plants (WPPs) are connected to the grid via long sea-cables characterized by their low X/R ratio and low short circuit ratio (SCR) \cite{connectionwindfarms}. The low SCR (and low inertia of the existing power grid) implies lower fault current contribution from the grid, and thus the grid connection is termed as a weak-grid. Small-signal stability studies have shown that low SCR weak-grid connections can introduce an open-loop zero in the right half plane of the system leading to oscillations which are poorly-damped or undamped \cite{7944679}. The low X/R ratio also enhances the P-V and Q-f coupling which otherwise would be negligible in a grid with high X/R ratio since the resistance is negligible in comparison to the reactance \cite{connectionwindfarms}. In a weakly-connected offshore WPP, a reactive power support device at the point of common coupling (PCC) helps to supply higher fault currents \cite{kundur2017power}, provide better reactive power support, and aid in post-fault voltage recovery \cite{9395086}. This further aids the WPP to operate smoothly during steady state as well as during fault-ride through (FRT.)

Synchronous condensers (SCs) have been known to stabilize power systems by providing damping to generator power swings and reactive power injection \cite{898204}. There has been some examples of stability enhancement of offshore WPPs with the help of a reactive power support device such as STATCOM, static var compensator (SVC,) or SC. e.g. A WPP connected to Texas' ERCOT grid experienced low frequency oscillations due to weak-grid connections which was subsequently seen to be damped by SVCs and SCs \cite{6344713}. SCs can also help control the wind-farm transient voltage under communication failure by using an `online sequential extreme learning machine' based voltage prediction method \cite{9395086}.

In a comparative analysis between STATCOM and SCs for wind-farm applications, results presented by \cite{9449775} show that a SCs can improve the stability of a WPP with GFL type-IV wind turbines (WTs) connected to a weak-grid; it is seen to improve the system stability even when it is not injecting any reactive power. On the other hand a STATCOM improves the stability of a WPP only by injecting reactive power.

Previous work supports the use of SCs in modern power systems as they increase the short-circuit performance, overload capabilities, inertial support, and also facilitate the integration of more inverter based resources \cite{9637902}. Weak-grid applications of SCs have been proposed along with battery energy storage system (BESS) in \cite{8905459} where the SC provided a significant improvement in overload capacity while retaining the fast response of a converter-integrated BESS. A similar application could be found for offshore WPPs with weak-grid connections. Although SCs can enhance the protection performance of a type-IV WPP, it has been seen that the converter control strategy also highly affects the overall performance of the WPP \cite{8357471}.

The key contributions of this paper are summarized as:
\begin{itemize}
    \item Small-signal stability quantification of an offshore WPP with the addition of SCs.
    \item Quantification of SCR enhancement of an offshore WPP with the addition of SCs.
\end{itemize}

The paper layout is as follows: Section \ref{sec: methodology} describes the system setup, its modelling (time-domain and frequency-domain small-signal model development,) and adopted methodology (small-signal stability analysis and time-domain fault simulation.) Section \ref{sec: SCR Enhancement} describes the equivalent SCR formulation including the effect of synchronous condenser. Section \ref{sec: Results and Discussion} provides the results obtained and a detailed analysis and discussion on the results. Finally, Section \ref{sec: Conclusion} summarizes the findings and provides a conclusion based on the studies presented in the paper.

\section{System Setup and Modelling}\label{sec: methodology}
\begin{figure*}[htbp]
    \centering
    \includegraphics[width=0.95\textwidth]{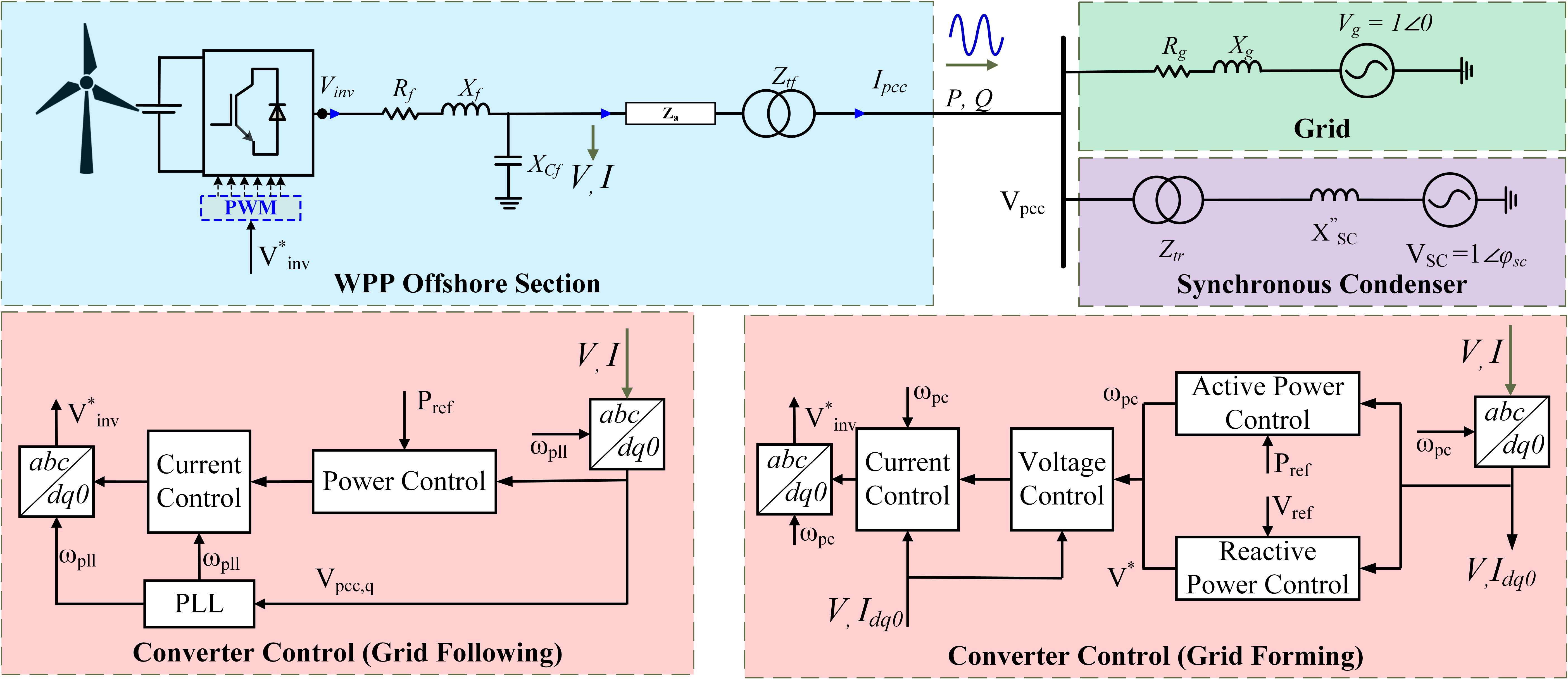}
    \caption{Aggregated WPP layout with synchronous condenser connected at PCC.}
    \label{fig: SC WPP}
\end{figure*}

The system under study consists of an aggregated WPP connected to a grid through an array cable and a transformer as shown in figure \ref{fig: SC WPP}. The array cable has an impedance $Z_{a}$, and the grid has an impedance $Z_{g} = R_g + jX_g$. A synchronous condenser with sub-transient reactance $Z_{sc} = jX^{"}_{sc}$ is connected at the PCC.

All parameters are represented in pu, and differential and algebraic variables are represented in small-signal pu when denoted by a lowercase letter, and are in large-signal SI units when denoted by an uppercase letter.

\subsection{Grid and Synchronous Condenser Model}
The grid is modelled as a Thevenin equivalent source, i.e. an ideal voltage source ($\Tilde{V}_g = 1\angle0^\circ$) behind an impedance $Z_g$. The load-flow behavior of a grid thus modelled would be that of a $P-\delta$ bus. The mathematical model of the grid used to build its time-domain model can be written as:
\begin{eqnarray}
    \Tilde{V}_g &=& 1\angle0^\circ\\
    \Tilde{I}_{g} &=& \frac{\Tilde{V}_g - \Tilde{V}_{pcc}}{R_{g} + jX_{g}}\\
    S_{g} &=& P_{g} + jQ_{g} = \Tilde{V}_g\Tilde{I}_{g}^*
\end{eqnarray}
where, $\Tilde{V}_g$ is the grid voltage, $\Tilde{V}_{pcc}$ is the PCC voltage, $Z_{g} = R_{g} + jX_{g}$ is the grid impedance, $\Tilde{I}_{g}$ is the current flowing from the grid towards PCC, and $S_{g} = P_{g} + jQ_{g}$ is the grid apparent power whose real part $P_{g}$ is the active power and imaginary part $Q_{g}$ is the reactive power. The $\Tilde{(\cdot)}$ above the variables indicate that they are phasor variables.

The state equation relating to the small-signal model of the grid is,
\begin{equation}
    L_{g}\Dot{i}_{g, dq} = -R_{g}i_{g, dq} + jX_{g}i_{g, dq} + v_{g,dq} - v_{pcc,dq}
\end{equation}

Similar to the grid, a synchronous condenser is also modelled as a Thevenin equivalent model, i.e. an ideal voltage source behind the sub-transient reactance. This modelling approach is valid for small-signal stability and fault studies since during the fault cases and fast transients, the sub-transient reactance of the synchronous condenser becomes active \cite{saadat2010power, miller1982reactive}. This approach has been adopted in the academic literature as well \cite{8443358}.

Synchronous condenser controls are bypassed for this study as their responses are very slow and only impact on the low-frequency ranges. The load-flow behavior of the synchronous condenser is that of a $P-V$ bus with a small active power loss $|P|<<1$ pu. The mathematical model of the synchronous condenser is thus given as,
\begin{eqnarray}
    \Tilde{V}_{sc} &=& 1\angle\varphi_{sc}\\
    \Tilde{I}_{sc} &=& \frac{\Tilde{V}_{sc} - \Tilde{V}_{pcc}}{jX_{sc}^" + R_{tr} + jX_{tr}}\\
    S_{sc} &=& P_{sc} + jQ_{sc} = \Tilde{V}_{sc}\Tilde{I}_{sc}^* \approx Q_{sc}
\end{eqnarray}
where $\Tilde{V}_{sc}$ is the synchronous condenser voltage, $\Tilde{I}_{sc}$ the current, $X_{sc}^"$ the sub-transient reactance, and $Q_{sc}$ is its reactive power contribution. The transformer is also modelled inside of the aforementioned equations and $R_{tr} + jX_{tr}$ represent the transformer impedance. The transformer used in this paper is a $Y-Y$ transformer, thus no phase-shift adjustment is necessary.

Now, the state equation relating to the small-signal model of the synchronous condenser is written as,
\begin{equation}
    \frac{X_{sc}^"}{\omega_0}\Dot{i}_{sc, dq} = R_{tr}i_{sc,dq} + j(X_{sc}^"+X_{tr})i_{sc, dq} + v_{sc, dq} - v_{pcc, dq}
\end{equation}

\begin{figure}[htbp]
    \centering
    \includegraphics[width = 0.4\textwidth]{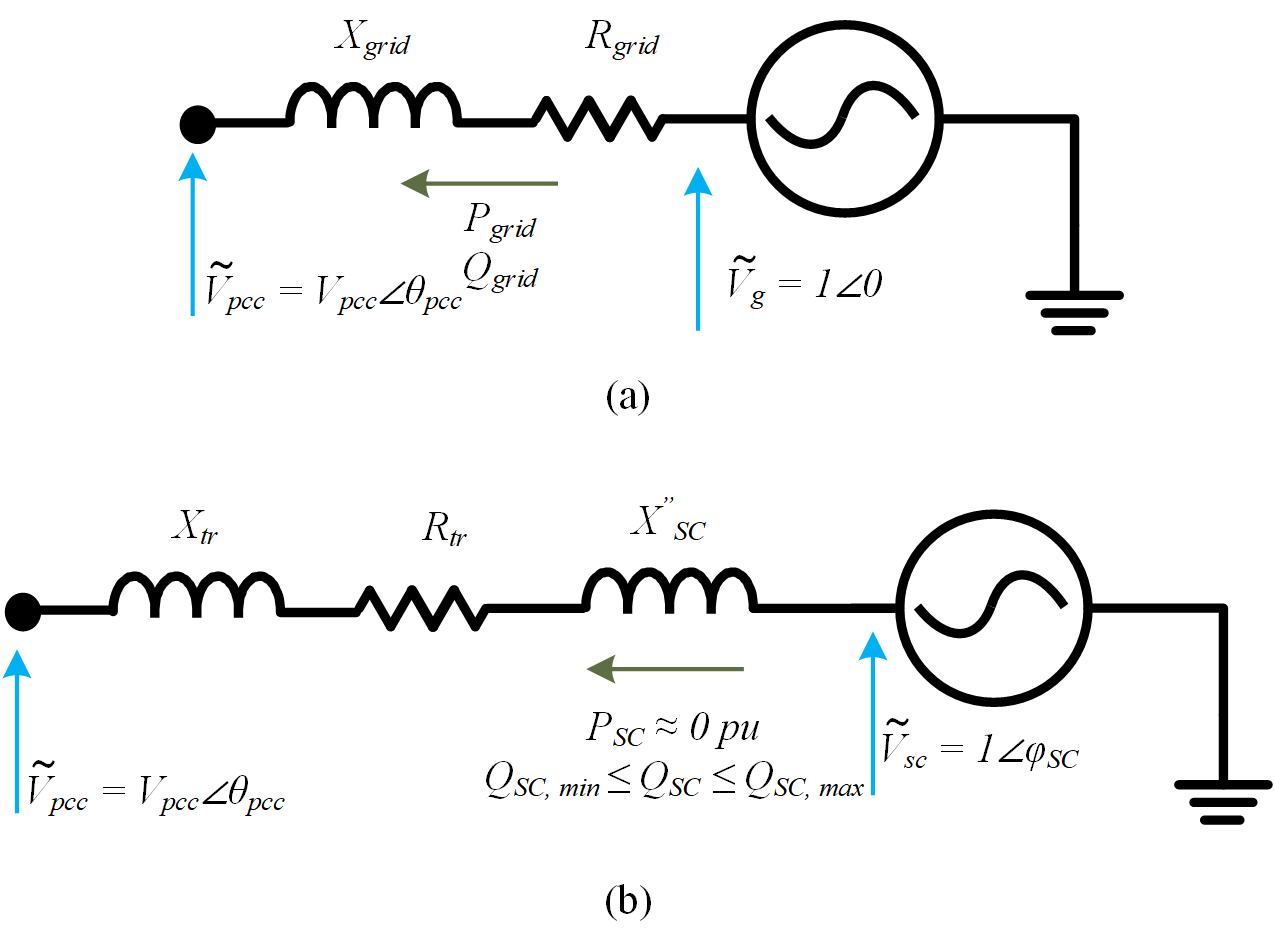}
    \caption{Thevenin equivalent models for (a) power grid, and (b) synchronous condenser and transformer used in the study.}
    \label{fig: SC Grid}
\end{figure}

The equivalent circuit of the power grid and synchronous condenser thus modelled are shown in Fig. \ref{fig: SC Grid}.

\subsection{WPP Converter Models}
Two different control methods for the WPP converters are selected for the studies, namely grid-following (GFL) control and grid-forming (GFM) control as shown in Figure \ref{fig: SC WPP}. Although the detailed modelling of these controls are not included, an averaged state-space model is presented in this paper.

\subsubsection{Grid-Following Control (GFL)}
The grid-following (GFL) control includes a phase-locked loop (PLL) as the synchronizing unit which drives the power and current controllers. The converter control model used is derived from \cite{ghosh2023improved} and readers are requested to refer to the same for details.

The linearized state-space equations of the GFL converter can be summarized as:
\begin{eqnarray}
    \Dot{\theta}_{pll} &=& \omega_0 + K_p^{(pll)}v_q + K_i^{(pll)}\mathcal{S}\\
    \Dot{\mathcal{S}} &=& v_q\\
    \Dot{\gamma} &=& -\frac{K_i^{(pc)}}{K_p^{(pc)}}\gamma + \frac{1}{K_p^{(pc)}}i_{dq}^*\\
    \Dot{\mathcal{O}}_{dq} &=& i_{dq}^* - i_{dq}
\end{eqnarray}
where, $\theta_{pll}$ is the PLL phase angle, $\omega_0=2\pi f_0$ is the nominal angular frequency, and $K_p^{(pll)}$ and $K_i^{(pll)}$ are the PLL proportional and integral gains respectively. $\gamma$ is the power controller state given as the integral of small-signal power mismatch, i.e. $\Dot{\gamma} =p^* - p_{pc}$, $K_p^{(pc)}$ and $K_i^{(pc)}$ are the PI-gains of the power controller. $p_{pc}$ and $q_{pc}$ are the small-signal active and reactive powers of the converter, $v_{dq}$ and $i_{dq}$ represent the voltage and current at the measurement points, $\mathcal{S}$ is a PLL-state which is given as $\Dot{\mathcal{S}} = v_q$, and $\mathcal{O}_{dq}$ are the current controller states. Now, the converter voltage reference signals can be written as:
\begin{equation}
    v_{idq}^* = v_{dq}^{(pcc)} + K_p^{(cc)}(i_{dq}^* - i_{dq}) + K_i^{(cc)}\mathcal{O}_{dq}
\end{equation}

\subsubsection{Grid-Forming Control (GFM)}
A virtual synchronous machine (VSM) based GFM control is implemented. For detailed model of the GFM converter control used in this paper, readers are requested to refer to \cite{Ghimire2023GridForming}.

The linearized state-space equations of the GFM converter are given as:
\begin{eqnarray}
    \Dot{\theta}_{pc} &=& \omega_{pc}\\
    J\Dot{\omega}_{pc} &=& p^* - p_{pc} - D_p\omega_{pc}\\
    \Dot{\mathcal{M}}_{dq} &=& v_{dq}^* - v_{dq}\\
    \Dot{\mathcal{O}}_{dq} &=& i_{dq}^* - i_{dq}
\end{eqnarray}
where, $J$ is the VSM inertia constant, $D_p$ is the VSM damping term, $\theta_{pc}$ is the angle given by the power controller, and $\mathcal{M}_{dq}$ and $\mathcal{O}_{dq}$ are the voltage controller and current controller states respectively. The voltage controller and current controller output equations are written as follows respectively.
\begin{eqnarray}
    i_{dq}^* &=& i_{dq}^{(pcc)} + G_{piv}(s)(v_{dq}^*-v_{dq}) + \frac{jv_{dq}}{X_{C_f}}\\
    v_{idq}^* &=& v_{dq}^{(pcc)} + G_{pic}(s)(i_{dq}^*-i_{dq}) + jX_fi_{dq}
\end{eqnarray}

\subsubsection{Filter and Array Cable Model}
Based on Kirchhoff's voltage and current laws, the state-space model of the converter's filter and array cable can be written as a set of differential-algebraic equations and are given as:
\begin{eqnarray}
    L_f\Dot{i}_{f,dq} &=& -R_fi_{f,dq} + jX_fi_{f,dq} - v_{c,dq} + v_{inv,dq}\\
    C_f\Dot{v}_{c,dq} &=& i_{f,dq} + j\omega_0C_fv_{c,dq} - i_{a,dq}\\
    L_{a,tf}\Dot{i}_{a,dq} &=& v_{c,dq} - R_{a,tf}i_{a,dq} + jX_{a,tf}i_{a,dq} - v_{pcc,dq}
\end{eqnarray}

Here, $R_f,\ L_f,\ C_f$ are the filter resistance, inductance, and capacitance, $L_a,\ R_a$ are the array cable inductance and resistance, and $L_{tf},\ R_{tf}$ are the transformer inductance and resistance respectively. Also, $R_{a,tf}$ and $L_{a,tf}$ are given as $R_a + R_{tf}$ and $L_{a} + L_{tf}$ respectively and $X_{a,tf} = \omega_0L_{a,tf}$.

\section{SCR Enhancement}\label{sec: SCR Enhancement}
The short-circuit ratio of a WPP at any given point is given as the ratio of the grid's fault MVA contribution to the WPP rated power.
\begin{equation}
    \text{i.e. }SCR = \frac{S_{g, f}}{S_{WPP}}
\end{equation}
where, $S_{g,f}$ is the grid fault contribution and $S_{WPP}$ is the WPP's rated apparent power.

Assuming $1$ pu voltage being maintained at the grid during the fault, and $S_{WPP}=1$ pu, we can rewrite the system SCR as:
\begin{equation}
    SCR = \frac{V_{g,pu}^2/|Z_{g,pu}|}{1\ pu} = \frac{1}{|Z_{g,pu}|}
    \label{eq: SCR basic}
\end{equation}

Assuming 1 pu voltage behind the sub-transient reactance, a synchronous condenser can be aggregated with the grid to observe an effective rise in the equivalent short circuit ratio (ESCR) at PCC.

From \eqref{eq: SCR basic}, we know that the SCR of a WPP at PCC can be given as the reciprocal of the pu grid impedance at $1$ pu grid voltage. In a WPP, the base SCR value at WT MV terminals without considering the synchronous condenser is be written as:
\begin{equation}
    \text{i.e. } SCR_o = \frac{1}{Z_{g,pu} + Z_{a,tf,pu}}
\end{equation}

Here, the impedance $Z_{g,pu}:= |Z_{g,pu}|$ and $Z_{a,tf,pu}:= |Z_{a,pu}+Z_{tf,pu}|$ are the absolute value of the sum of the respective impedances. The same convention is followed for all other impedances henceforth unless otherwise stated.

Considering the stabilizing effect and fault MVA contribution of a synchronous condenser on SCR evaluation, we can view the grid and synchronous condenser pair from PCC as two voltage sources in parallel. Thus we can write for the pu impedance seen at PCC as: 
\begin{equation}
    Z_{PCC, pu} = \frac{Z_{g,pu}\cdot Z_{sc,pu}}{ Z_{g,pu}+Z_{sc,pu}} <\min{(Z_{g,pu},Z_{sc,pu})}
\end{equation}

The effective SCR at WT MV terminal can now be written as:
\begin{equation}
    SCR_{sc} = \frac{1}{Z_{PCC, pu} + Z_{a,tf,pu}}
\end{equation}

Here, since $Z_{PCC, pu}<Z_{g,pu}$, it implies that SCR$_{sc}>SCR_o$. Thus effective SCR at the WT MV terminal considering the synchronous condenser is higher than that without the synchronous condenser. The physical interpretation of this is that the short-circuit MVA contribution of the synchronous condenser increases the overall fault MVA of the system, thus increasing the effective SCR.

We can now write for this new effective SCR including the effect of the synchronous condenser as:
\begin{equation}
    SCR_{sc}  = \frac{Z_{g,pu}+Z_{sc,pu}}{ Z_{g,pu}\cdot Z_{sc,pu} + Z_{a,tf,pu}(Z_{g,pu}+Z_{sc,pu})}
    \label{eq: SCR enhanced}
\end{equation}

The SCR enhancement quantification provided here in equation \ref{eq: SCR enhanced} is verified with SCR values calculated from time-domain fault simulations including the effect of SC in the system.

\section{Results and Discussion}\label{sec: Results and Discussion}
The small-signal and time-domain simulation models developed for the test system presented in Fig. \ref{fig: SC WPP} are subjected to different tests. Frequency domain studies involving the step-responses of power and voltage controllers of the converter controls are presented in this section where small-signal stability information are observed, e.g. controller reference tracking, oscillations and damping based on step responses, and stability/instability based on eigenvalue plots. The WPP's SCR is calculated before and after the addition of the synchronous condenser based on (a) fault simulations and (b) theoretical calculations based on the method described in Section \ref{sec: SCR Enhancement}, and are presented in this section.

\subsection{Small Signal Stability Analysis}
For the frequency-domain analysis and studies, different grid conditions and operating points are chosen. IEEE guidelines \cite{653230} classify system strength in terms of SCR: a very weak grid has very low SCR$<2$, a weak grid has $2<SCR\leq 3$, and a strong grid has SCR$\geq3$. An offshore WPP's SCR is measured from the turbine MV terminals. SCR has a critical effect on system stability and study for extreme SCR cases become necessary for such systems. In order to study the effect of synchronous condensers on stability of a WPP for different grid strengths, three grid strength cases are chosen and summarized in Table \ref{tab: Grid Cases} in terms of system SCR and X/R ratio.
\begin{table}[htbp]
    \centering
    \caption{Grid cases in terms of SCR at WT MV terminal.}
    \label{tab: Grid Cases}
    \begin{tabular}{ccc}
    \hline
        Grid Case & Base-Case SCR & X/R Ratio\\
        \hline\hline
        Weak Grid & 1.6 & 5 \\
        Normal Grid & 3.2 & 14.8\\
        Strong Grid & 4.12 & 14.8\\\hline
    \end{tabular}
\end{table}

\begin{figure}[htbp]
    \centering
    \includegraphics[width = 0.5\textwidth, trim = {0.75cm, 0.5cm, 1cm, 0cm}, clip]{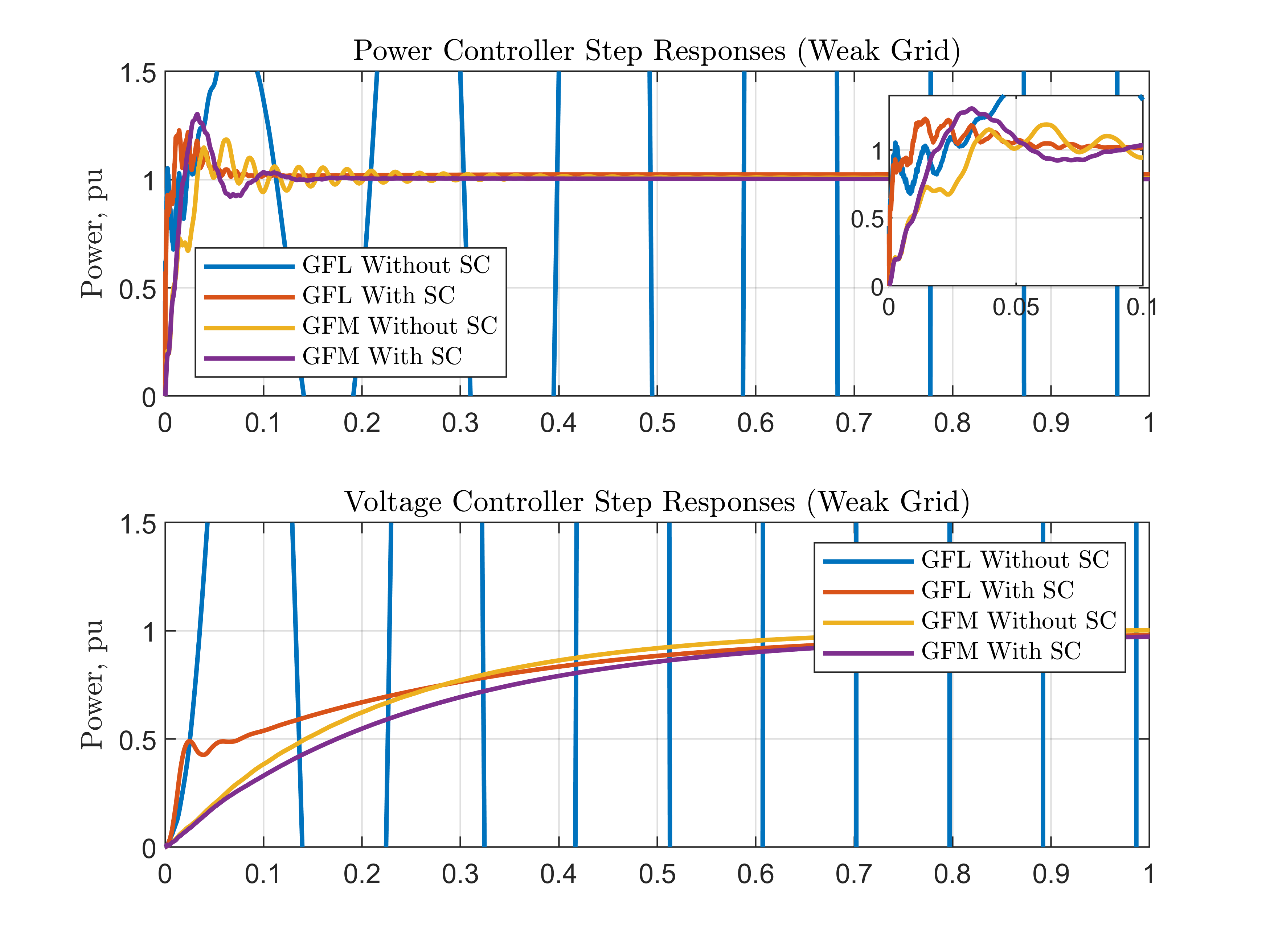}
    \caption{Power and voltage controller step response of GFL and GFM converters in a weakly connected WPP with and without synchronous condenser.}
    \label{fig: PV Step WeakGrid}
\end{figure}

\begin{figure}[htbp]
    \centering
    \includegraphics[width = 0.5\textwidth, trim = {0.75cm, 0.5cm, 1cm, 0cm}, clip]{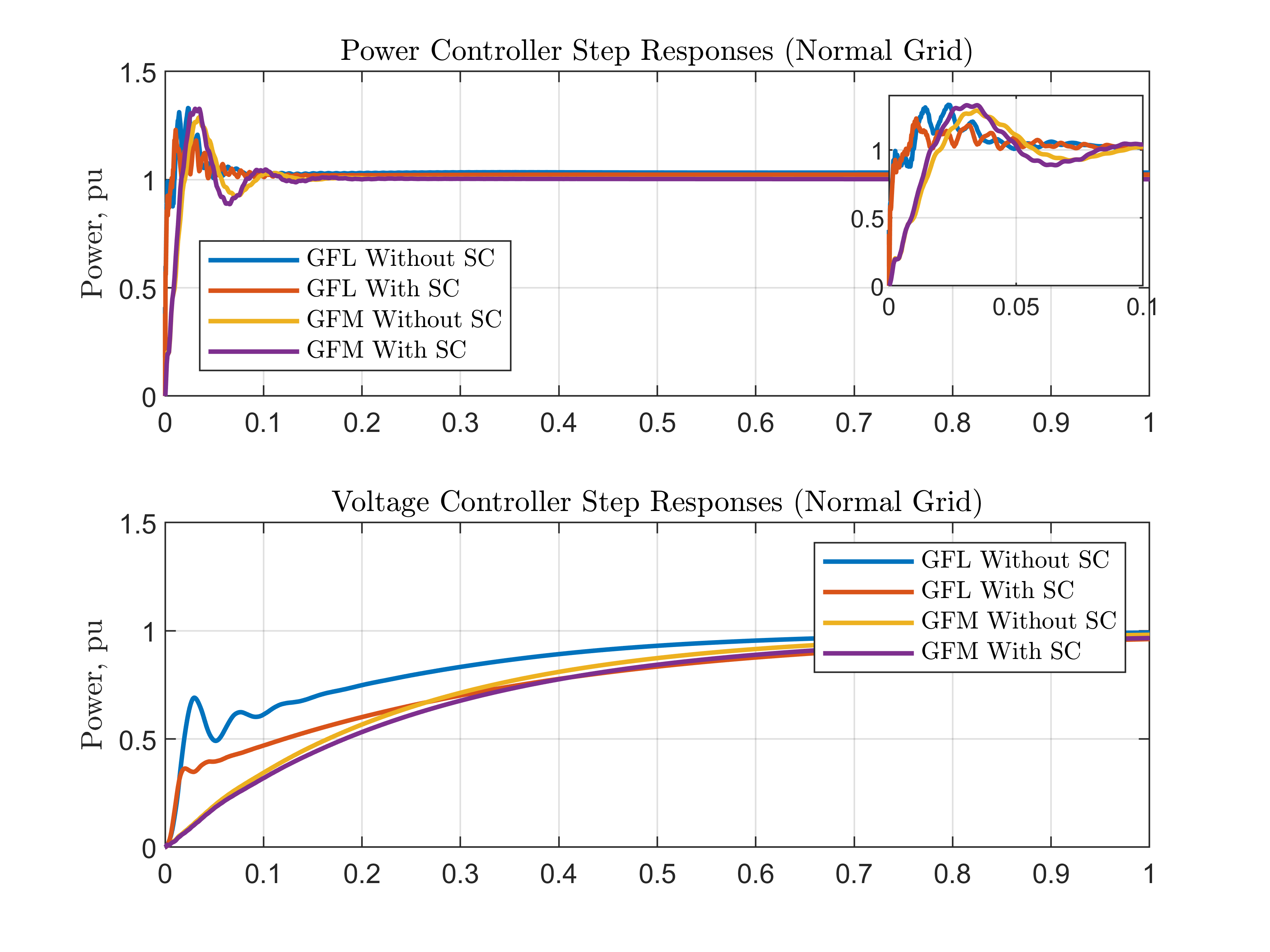}
    \caption{Power and voltage controller step response of GFL and GFM converters in a normal-grid connected WPP with and without synchronous condenser.}
    \label{fig: PV Step NormalGrid}
\end{figure}

\begin{figure}[htbp]
    \centering
    \includegraphics[width = 0.5\textwidth, trim = {0.75cm, 0.5cm, 1cm, 0cm}, clip]{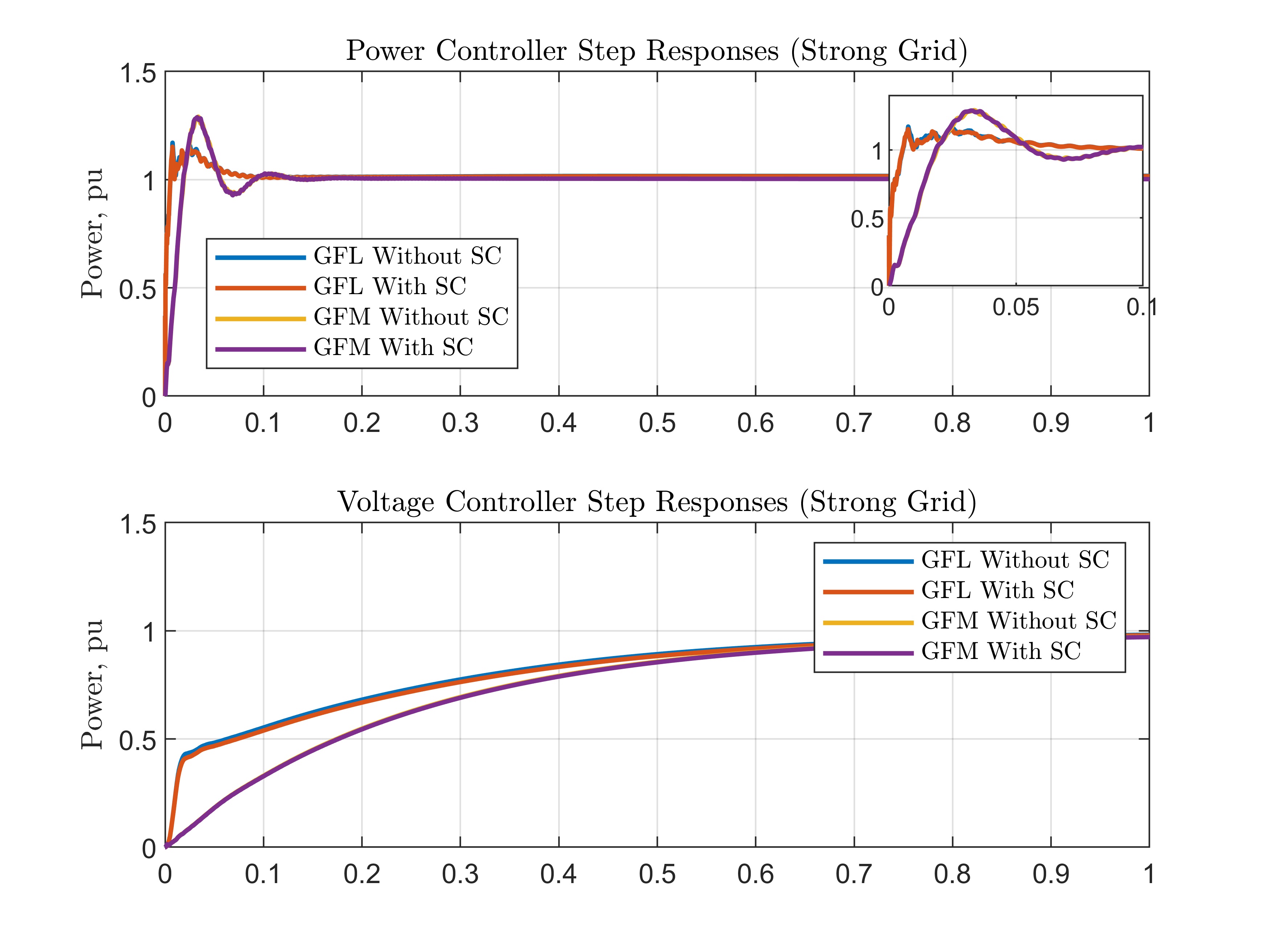}
    \caption{Power and voltage controller step response of GFL and GFM converters in a strongly connected WPP with and without synchronous condenser.}
    \label{fig: PV Step StrongGrid}
\end{figure}

The step responses of the power and voltage controllers for each of these cases are presented in Fig. \ref{fig: PV Step WeakGrid} (weak grid,) Fig. \ref{fig: PV Step NormalGrid} (normal grid,) and Fig. \ref{fig: PV Step StrongGrid} (strong grid.) System references/operating points for the presented step responses are grid reference voltage $V_g^*=1$ pu, turbine reference voltage $V_{turb}^*=1$ pu, and turbine reference power $P_{turb}^*=1$ pu respectively.

The step responses in Fig. \ref{fig: PV Step WeakGrid} show that for the provided system responses in a weakly connected WPP with GFL control, adding a synchronous condenser at PCC can stabilize the previously unstable system. Furthermore, although instability is not observed in the response of GFM control, the inset in the power controller step responses shows that the initial oscillations are damped with the addition of synchronous condenser.

For the given operating point in a normal grid connected GFM WPP, Fig. \ref{fig: PV Step NormalGrid} shows that no instability issue is seen in the base case and there is no significant effect on power and voltage controller step responses after the addition of the synchronous condenser. The inset on the power controller step response plot also clarifies that apart from minor differences in the transient behavior of the controller, there is no major impact on system performance. The voltage controller response shown in Fig. \ref{fig: PV Step NormalGrid} shows that synchronous condenser has made the voltage controller response slower. However, no further noticeable difference on dynamic or steady-state performance is seen to have been introduced by the synchronous condenser, suggesting that synchronous condenser can enhance the system damping.

The system is further tested on a high SCR grid (strong-grid case with SCR$=4.12$.) The step plots shown in \ref{fig: PV Step StrongGrid} show that in a strongly connected WPP, addition of synchronous condenser doesn't change the system stability and small-signal performance in any significant way.

To better understand the impact of synchronous condensers on WPP stability, further analysis are presented in terms of system eigenvalues for different operating points. For each of the sub-cases in Table \ref{tab: Grid Cases}, eigenvalue analysis are performed for a range of 27 different operating points defined as grid voltage reference($V_g^*$,) wind turbine voltage reference ($V_{turb}^*$,) and wind turbine power references ($P_{turb}^*$) as shown in Table \ref{tab: Operating Points}.
\begin{table}[htbp]
    \centering
    \caption{Operating Points}
    \label{tab: Operating Points}
    \begin{tabular}{cc}
    \hline
        Reference & Values\\
        \hline\hline
        Grid Voltage Reference ($V_{g}^*$) & $\begin{bmatrix}0.92 & 1.0 & 1.08
        \end{bmatrix}$\\
        Turbine Voltage Reference ($V_{turb}^*$) & $\begin{bmatrix}0.92 & 1.0 & 1.08
        \end{bmatrix}$\\
        Turbine Power Reference ($P_{turb}^*$) &  $\begin{bmatrix}0.1 & 0.5 & 1.0
        \end{bmatrix}$\\\hline
    \end{tabular}
\end{table}

\begin{figure}[htbp]
     \centering
     \includegraphics[width=0.45\textwidth, trim = {0.75cm, 0.5cm, 1cm, 0cm}, clip]{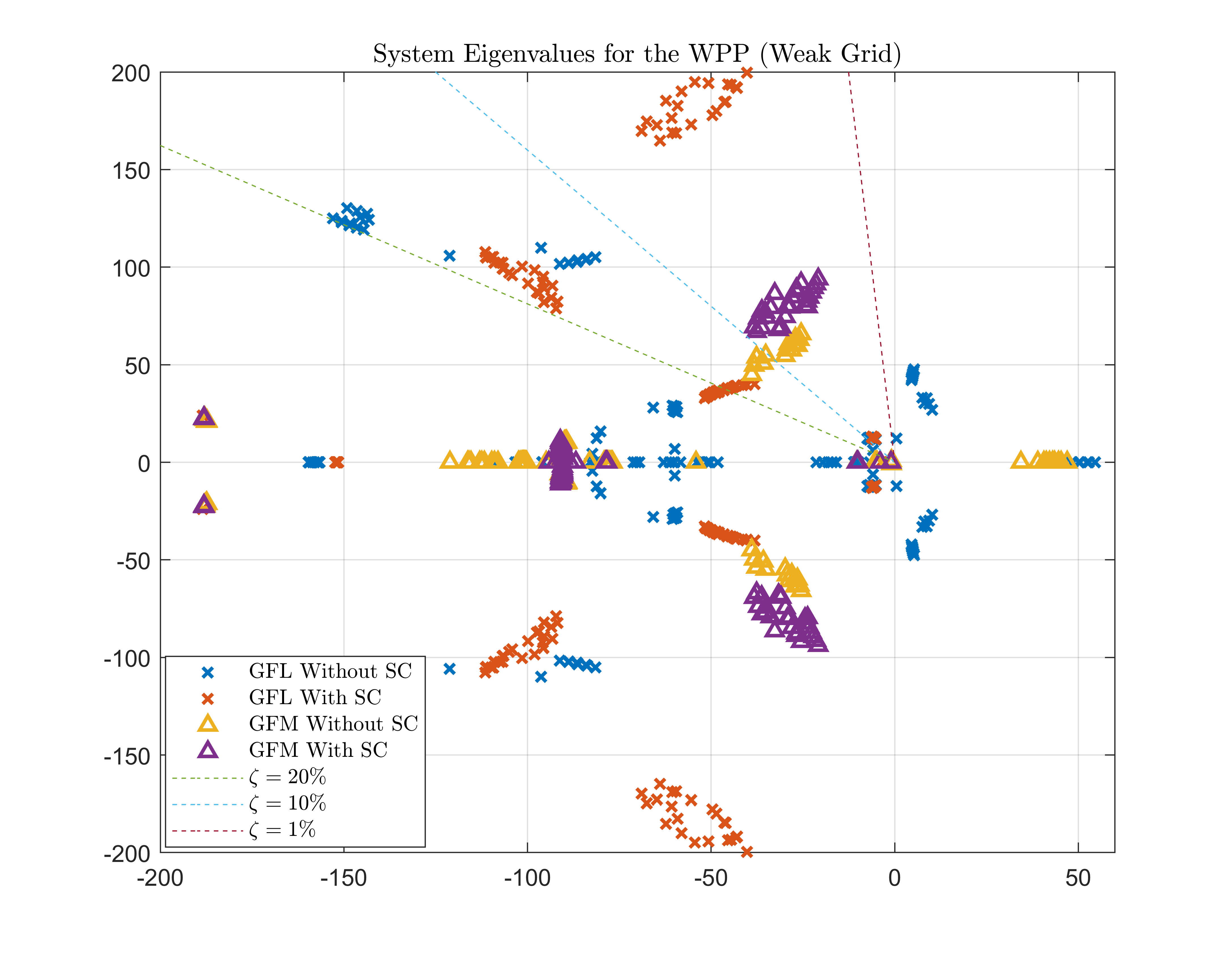}
     \caption{Eigenvalues for weak-grid case (27 standard operating points superimposed)}
     \label{fig: Eigval WkGrid 27}
\end{figure}

\begin{figure}[htbp]
     \centering
     \includegraphics[width=0.45\textwidth, trim = {0.75cm, 0.5cm, 1cm, 0cm}, clip]{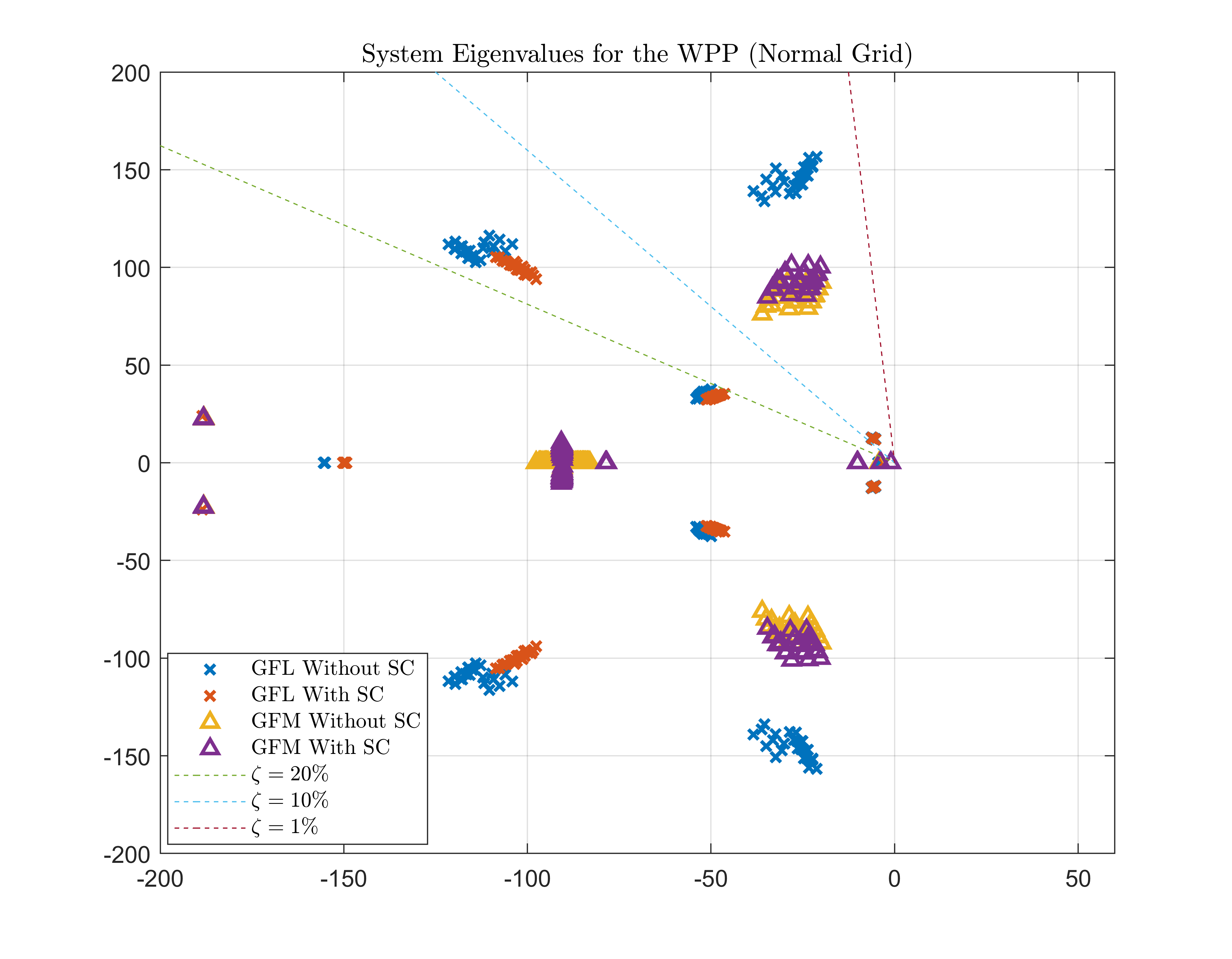}
     \caption{Eigenvalues for normal-grid case (27 standard operating points superimposed)}
     \label{fig: Eigval StrGrid 27}
\end{figure}

\begin{figure}[htbp]
     \centering
     \includegraphics[width=0.45\textwidth, trim = {0.75cm, 0.5cm, 1cm, 0cm}, clip]{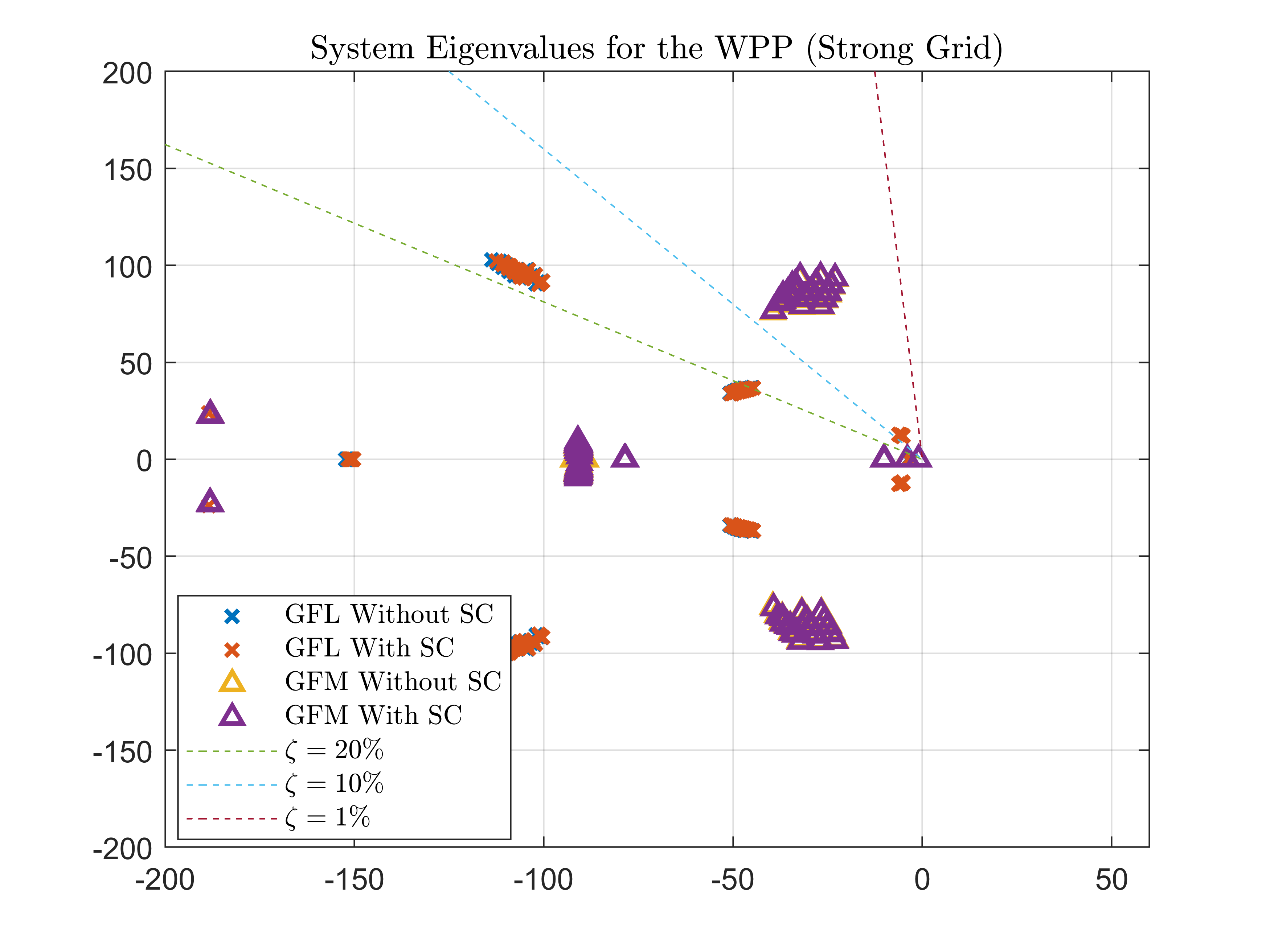}
     \caption{Eigenvalues for Strong-grid case (27 standard operating points superimposed)}
     \label{fig: Eigval VStrGrid 27}
\end{figure}

The eigenvalues of the system for the 27 operating points given in \ref{tab: Operating Points} are coalesced in one figure and plotted in Fig. \ref{fig: Eigval WkGrid 27} for weak-grid case, Fig. \ref{fig: Eigval StrGrid 27} for normal-grid case, and \ref{fig: Eigval VStrGrid 27}. Eigenvalue damping is also defined and marked in the eigenvalue plots with dotted lines. The eigenvalue damping is given as:
\begin{equation}
    \zeta_i = -\frac{\Re\{\lambda_i\}}{|\lambda_i|}
\end{equation}
where $\lambda_i$ is the $i^{th}$ eigenvalue. Eigenvalue damping ratio are marked in the plots are helpful for understanding system stability as low and near-synchronous frequency oscillations with higher than $20\%$ damping ratio ensure well-damped oscillations during normal operating condition.

The poles on the right hand side of the $s-$plane in figure \ref{fig: Eigval WkGrid 27} suggest that irrespective of GFL or GFM control, for operating points with low grid and turbine voltage references ($V_{turb}^* = 0.92$ pu, $V_g^* = 0.92$ pu,) the weakly connected offshore WPP exhibits small-signal instability. For the nominal operating point, the step responses in Fig. \ref{fig: PV Step WeakGrid} showed that the GFM WPP system stayed stable even without the addition of the synchronous condenser. However, for other operating points, as seen from Fig. \ref{fig: Eigval WkGrid 27}, the system goes unstable and it is stabilized by the addition of synchronous condenser at the PCC. Near-synchronous frequency modes with poor damping (i.e. poles with $\zeta<20\%)$ are also seen in cases where turbine and grid voltage references were at their lower levels, i.e. $V_{turb}^* = 0.92$ pu, $V_g^* = 0.92$ pu.

The normal-grid connected case for the offshore WPP did not exhibit any stability issues with or without the synchronous condenser across the selected 27 operating points. There are some changes in the eigenvalues location as shown in Fig. \ref{fig: Eigval StrGrid 27}, however, the eigenvalue damping and frequencies change only slightly and there is no significant change in system stability. This was also observed in the step responses calculated for the nominal case in Fig. \ref{fig: PV Step NormalGrid}. Further, the low-damped, near-synchronous eigenvalues from weak-grid case are seen to be well-damped in this case. This further suggests that some unstable system modes are stabilized with synchronous condensers in a weakly connected system, while a normal-grid connection do not necessarily require synchronous condensers to enhance system stability. Furthermore, poorly damped near-synchronous oscillatory modes (not prevalent in normal grids) are not strongly affected by synchronous condensers in weak grids; grid strength play a crucial role for the elimination/damping of near-synchronous modes than synchronous condensers do.

Further eigenvalue analysis for a strong-grid with SCR$ = 4.12$ is performed on all the predefined operating points (see Table \ref{tab: Operating Points}.) The eigenvalue plots for this case presented in figure \ref{fig: Eigval VStrGrid 27} also show that for a strongly connected offshore WPP, adding a synchronous condenser does not change the system poles in any discernible way.

\subsection{Equivalent Short Circuit Ratio}
\begin{figure}[htbp]
    \centering
    \includegraphics[width = 0.5\textwidth, trim = {1.75cm, 0.5cm, 1cm, 0cm}, clip]{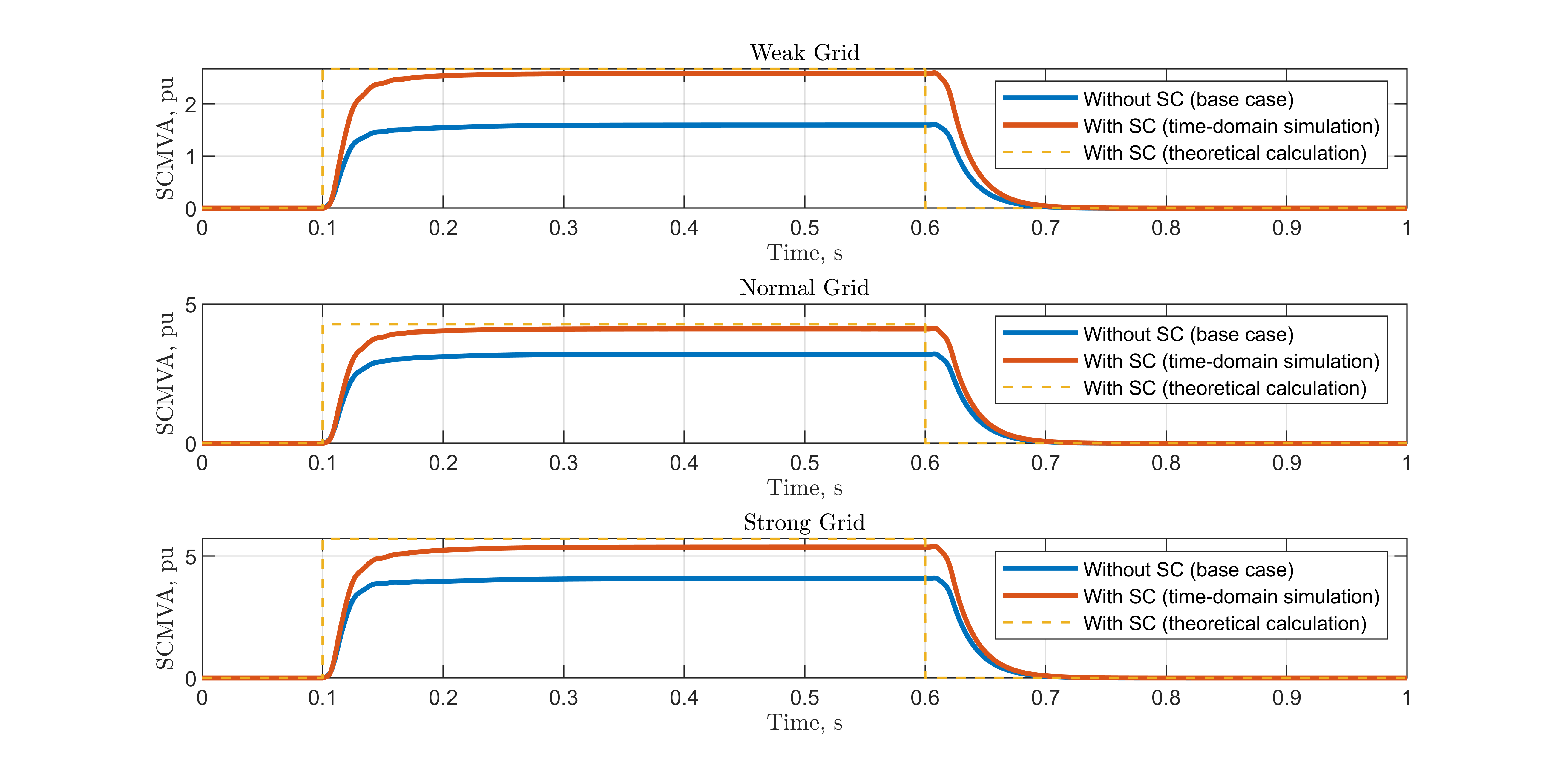}
    \caption{SCMVA levels based on theoretical calculation and time-domain simulations.}
    \label{fig: SCR Enhanced}
\end{figure}

The ESCR of the system before and after the addition of the synchronous condenser calculated using the time-domain simulation models as well as based on the procedure described in \ref{sec: SCR Enhancement}. Short-circuit levels for both these methods are presented in figure \ref{fig: SCR Enhanced}. The short-circuit MVA is presented in pu with the base of the WPP apparent power rating $S_{WPP}$. A summary of the SCR values based on these results are summarized in Table \ref{tab: SCR Enhanced RMS WT MV} below. In either of these cases, the contribution of converters on the effective SCR is not considered.

\begin{table}[htbp]
    \centering
    \caption{SCR enhancement of a WPP at WT MV terminals based on theoretical calculation vs fault MVA simulations: with and without synchronous condenser (SC.)}
    \label{tab: SCR Enhanced RMS WT MV}
    \begin{tabular}{cccc}
    \hline
    Grid Case & W/O SC & W/ SC (theoretical) & W/ SC (simulation)\\
    \hline\hline
         Weak Grid & SCR = 1.6 & SCR = 2.67 & SCR = 2.58\\
         Normal Grid & SCR = 3.2 & SCR = 4.28 & SCR = 4.11\\
         Strong Grid & SCR = 4.12 & SCR = 5.71 & SCR = 5.37\\
    \hline
    \end{tabular}
\end{table}

From \ref{fig: SCR Enhanced} and Table \ref{tab: SCR Enhanced RMS WT MV}, the effective SCR of the WPP is seen to have increased significantly after the introduction of the synchronous condenser. This also shows a good match between the theoretical calculation and time-domain simulation based results. Since converter control can interact with synchronous condenser control, further detailed information on the overall system stability needs to be gathered with a more detailed synchronous condenser model including its control dynamics and immitance-based stability analysis methods.

\section{Conclusion}\label{sec: Conclusion}
A state-space and time-domain model of the aggregated offshore WPP was constructed and the effect of addition of a synchronous condenser in such a system was studied for different cases. The analysis shows that for a weakly connected offshore WPP, synchronous condenser can have significant positive impact on stability. It can enhance the small-signal stability, improve the system response by damping out oscillations, and enhance the SCR of the system. The stabilizing effect is more significant for GFL WPPs than for GFM WPPs. For a normal-grid connected and strongly connected offshore WPP, no significant stability improvements were observed for the tested 27 operating points. However, the effective SCR of the system was seen to have been improved due to the addition of the synchronous condenser in both these cases. Even though frequency domain analysis didn't show significant improvement in small signal stability of the WPPs in these two cases, the synchronous condenser can still aid in the overall fault response of the WPP and improve the dynamic stability.

With GFM control of the turbines, the stability issues were observed for fewer number of operating points. Further analysis seems necessary to fully understand the stability performance of both synchronous condensers and GFM converters, and to investigate if a GFM converter can have same or better stability performance as a GFL converter and synchronous condenser pair. A stability region in parametric domain (e.g. SCR vs X/R) can also be plotted for system with and without synchronous condensers and the overlapping zones could be analyzed. 

Small-signal stability quantification was provided for offshore WPPs with both GFL and GFM control with the addition of synchronous condensers. Effect of synchronous condensers on the SCR and SCMVA level for such systems were defined, mathematically described, and validated with time-domain fault simulations.

\section{Legal Disclaimer}\label{sec: Acknowledgements and Disclaimer}
This work was supported by Innovation Fund Denmark under the project Ref. no. 0153-00256B. Figures and values presented in this paper should not be used to judge the performance of Siemens Gamesa Renewable Energy technology as they are solely presented for demonstration purpose. Any opinions or analysis contained in this paper are the opinions of the authors and not necessarily the same as those of Siemens Gamesa Renewable Energy.

\bibliography{bib/Bibliography}
\bibliographystyle{ieeetr}

\end{document}